\documentclass[twocolumn]{aastex701}
\usepackage{graphicx}
\usepackage{dcolumn}
\usepackage{bm}
\usepackage{amssymb}
\usepackage{amsmath}
\usepackage{soul, color, xcolor}
\usepackage{epstopdf}
\usepackage{booktabs}

\begin{document}

\title{Detecting and Identifying Supernovae in AGN Accretion Disks}

\author[orcid=0009-0007-0717-3667]{Zhi-Peng Ma}
\affiliation{Department of Astronomy, School of Physics, Huazhong University of Science and Technology, Wuhan, Hubei 430074, China}
\email{d202380093@hust.edu.cn}

\author[orcid=0000-0003-4976-4098]{Kai Wang}
\affiliation{Department of Astronomy, School of Physics, Huazhong University of Science and Technology, Wuhan, Hubei 430074, China}
\email[show]{kaiwang@hust.edu.cn}

\begin{abstract}

Stars may become embedded in active galactic nucleus (AGN) accretion disks and eventually explode as supernovae (SNe), whose detection would provide direct evidence for the existence and evolution of stars in AGN disks. We investigate the electromagnetic emission of such disk SNe using an analytic shock dynamics model coupled with a one-zone radiation model. Applying this framework to four progenitor types over a range of AGN black hole masses and Eddington ratios, we find that disk SNe around lower-mass black holes (BHs) or at lower Eddington ratios are more readily detectable against the underlying AGN emission and can remain detectable over a broader range of electromagnetic bands, from the infrared to soft X-rays. However, detectability against the AGN background does not guarantee that disk SNe can be distinguished from isolated SNe with CSM interaction. We therefore further compare disk SNe with isolated SNe with and without CSM interaction and find substantial degeneracies in their macroscopic observables.
Nevertheless, we identify distinctive signatures that are robust across progenitor types. The characteristic emission shifts from the EUV for \(M_{\rm BH}\lesssim10^7\,M_\odot\) toward the UV/optical for \(M_{\rm BH}=10^8\,M_\odot\), with the visibility timescale increasing from less than a day to \(\sim3\text{--}70\) days and \(T_{\rm peak}\sim3\text{--}8\) eV in the latter case. The UV/optical regime around \(10^8\,M_\odot\) BHs therefore offers particularly promising prospects for detecting disk SNe.

\end{abstract}



\section{Introduction} 
Active galactic nuclei (AGNs) provide a unique environment for stellar evolution and explosive transients. Their accretion disks can host stars formed in situ or captured from the surrounding nuclear stellar population, allowing massive stars to become embedded in the disk~\citep{artymowicz1993star,collin1999star,wang2011star,fabj2020aligning,cantiello2021stellar}. The dense gaseous environment, together with stellar migration and dynamical interactions, may facilitate stellar growth and ultimately trigger supernova explosions~\citep{dittmann2020star,dittmann2021accretion,grishin2021supernova}. The remnant compact objects, including black holes (BHs) and neutron stars (NSs), may remain embedded in the disk and interact dynamically, eventually forming compact-object binaries and undergoing mergers~\citep{cheng1999formation,bartos2017rapid,leigh2018rate,mckernan2020black}. Such mergers provide an additional channel for energetic transient events in AGN disks~\citep{kimura2021outflow,tagawa2023observable,tagawa2026properties}.

Both SNe and compact-object mergers in AGN disks can generate electromagnetic and high-energy neutrino emission through interactions between their ejecta or merger-driven outflows/jets and the surrounding disk material~\citep{grishin2021supernova,zhu2021high,tagawa2023shock,zhou2023high,ma2025electromagnetic}, while compact-object mergers additionally produce gravitational-wave (GW) signals~\citep{mckernan2020black,samsing2022agn}. These multimessenger signatures provide promising observational probes of the stellar-disk interaction scenario described above. Notably, the BBH merger event GW190521 reported by LIGO/Virgo was temporally associated with an optical flare observed $\sim30$ days after the merger~\citep{graham2020candidate}, while several nuclear flares have also been found to be statistically associated with LIGO/Virgo BBH merger events~\citep{graham2023light}. Such observations provide intriguing hints that some compact-object mergers may occur in AGN disks, motivating further investigation of the stellar-disk scenario.

Although the aforementioned BBH observations provide intriguing hints in favor of the stellar-disk scenario, they do not yet constitute compelling evidence~\citep{palmese2021ligo,veronesi2025agn}. Electromagnetic emission from SNe embedded in AGN disks, on the other hand, may offer a more direct and promising observational probe of this scenario. In particular, the expected rate of SN explosions may substantially exceed that of compact-object mergers, since SNe can arise from the deaths of individual massive stars, whereas compact-object mergers require the additional formation and subsequent coalescence of a compact-object binary. The volumetric rate of SNe in AGN disks has been estimated to reach $\mathcal{R}_{\rm SN}\lesssim100~{\rm Gpc^{-3},yr^{-1}}$ under optimistic assumptions about star formation and the AGN population~\citep{grishin2021supernova}, although the actual rate is subject to substantial uncertainties. Other studies have likewise considered potentially significant populations of stellar explosions embedded in AGN disks~\citep{moranchel2021supernova}. By comparison, theoretical estimates of the BBH merger rate in AGN disks typically span $\mathcal{R}_{\rm BBH}\sim0.1-20~{\rm Gpc^{-3},yr^{-1}}$, with substantial uncertainties depending on the AGN disk properties and embedded compact-object population~\citep{ford2022binary,arca2023quiescent,ishibashi2024gravitational}. This difference in the expected event rates suggests that electromagnetic searches for disk-SNe may provide a more promising and statistically accessible route to testing the stellar-disk scenario.accessible route to testing the stellar-disk scenario.

The interaction between supernova ejecta and circumstellar material (CSM) has been extensively studied both observationally and theoretically, providing a well-established framework for understanding the diverse electromagnetic signatures of interacting SNe~\citep{chevalier1982self,filippenko1997optical,moriya2013analytic,smith2016interacting}. Theoretical studies of electromagnetic emission from SNe embedded in AGN disks have also adopted both analytical and numerical approaches. \cite{grishin2021supernova} developed an analytical model for shock breakout and light curves, complemented by radiation-hydrodynamic calculations and 2D simulations, and found that the breakout can produce bright blue, UV, or X-ray flares. \cite{li2023core} and \cite{she2025anisotropy} further modeled the light curves of core-collapse SNe with different progenitors, finding that ejecta--disk interaction can produce prominent UV/optical emission. However, these treatments adopt simplified disk dynamics and geometry. In contrast, 3D hydrodynamic simulations have been used to study the non-spherical evolution of SN remnants and disk rotation~\citep{moranchel2021supernova,cook2026agn}, but do not calculate detailed electromagnetic light curves.

However, a systematic investigation of the observational prospects of disk-SNe is still lacking. In particular, it remains unclear which progenitor types are most favorable for detection, which AGN environments are most likely to host observable disk-SNe, and which observable features could distinguish them from SNe occurring in the host galaxy. In this work, we explore these questions by systematically examining the luminosity and characteristic temperature of disk SNe across different progenitors and AGN disk properties. We further identify potential observational signatures that are distinctive to the AGN-disk environment and discuss how such features may help reveal disk-SNe in future observations. Ultimately, detecting and identifying these events would provide a direct observational probe of the stellar population and explosive transients in AGN accretion disks.

This paper is organized as follows. Section~\ref{sec:model} describes the AGN disk model, shock dynamics, and one-zone radiation model used to calculate the SN light curves. In Section~\ref{sec:obs}, we apply this framework to four progenitor types across a range of AGN parameters and investigate the observability of disk SNe. Section~\ref{sec:dis} compares disk SNe with isolated SNe, with and without CSM interaction, to identify potentially distinguishing observational signatures. Finally, Section~\ref{sec:sum} summarizes our results and discusses their implications.

\section{Model description} 
\label{sec:model}
\subsection{Disk Structure}
In this work, we construct the AGN disk structure based on the SG model~\citep{sirko2003spectral}, which provides a physically motivated description of the radial disk structure in the inner region, extending up to $\lesssim 10^5R_{\rm g}$, where $R_{\rm g}=GM_{\rm BH}/c^2$ denotes the gravitational radius. We employ the publicly available Python package \texttt{pagn}~\citep{gangardt2024pagn}, which efficiently solves the disk structure equations for either the SG model or the TQM model~\citep{thompson2005radiation} given a set of input parameters. For our fiducial model, we adopt the standard SG disk parameters with Shakura--Sunyaev viscosity parameter $\alpha=0.01$, Eddington ratio $l_{\rm E}=0.5$, radiative efficiency for accreted matter $\epsilon=0.1$, and hydrogen abundance $X=0.7$~\citep{sirko2003spectral}. 
With these parameters, the AGN bolometric luminosity is determined by $L_{\rm bol}=l_{\rm E}L_{\rm Edd}$, where $L_{\rm Edd}=4\pi GM_{\rm BH}m_{\rm p}c/\sigma_{\rm T}$ is the Eddington luminosity. Eddington ratio $l_{\rm E}$ also determines the AGN background emissions. The mean molecular weight of the disk gas is calculated from the hydrogen abundance as $\mu=1/(2X+\frac{3}{4}(1-X))$.

The resulting radial profiles of the disk properties, including the mass density $\rho(r)$, scale height $H(r)$, temperature $T_{\rm d}(r)$, and opacity $\kappa(r)$, are obtained from the numerical solution. For the vertical structure, we assume the disk to be vertically isothermal at a given radius. Under vertical hydrostatic equilibrium, this assumption leads to a Gaussian density profile~\citep{grishin2021supernova},
\begin{equation}
T_{\rm d}(r,z)=T(r),
\end{equation}
and
\begin{equation}
\rho_{\rm d}(r,z)=\rho(r)\exp\left(-\frac{z^2}{2H^2}\right).
\end{equation}
\subsection{Shock Dynamics}
A SN explosion occurring at radius $r$ with ejecta energy $E_{\rm ej}$ and ejecta mass $M_{\rm ej}$ interacts with the surrounding disk gas and drives a shock wave. With the shock persistently sweeping up matter ahead of it, the ejecta kinetic energy is converted into the thermal energy of the shocked material, which can subsequently be radiated through shock breakout and photon diffusion processes. Therefore, the shock velocity evolution is a key factor in determining the electromagnetic emission. In general, the SN ejecta density profile can be described by a broken power-law distribution with different indices. The inner ejecta follows $\rho_{\rm ej}\propto R^{-d}$, where $d\approx0$, while the outer ejecta follows $\rho_{\rm ej}\propto R^{-n}$, with $n\simeq7$--$12$ depending on the progenitor structure~\citep{matzner1999expulsion,truelove1999evolution}. The transition between the two regions occurs at the characteristic velocity $v_{\rm t}=\sqrt{\frac{2(5-d)(n-5)E_{\rm ej}}{(3-d)(n-3)M_{\rm ej}}}$. Since the outer ejecta expands at higher velocities, we define the maximum ejecta velocity as $v_{\rm max}=2v_{\rm t}$ in this work. For ejecta with an outer density profile $\rho_{\rm ej}\propto R^{-n}$ expanding into an ambient medium with a power-law density profile $\rho_{\rm gas}\propto R^{-s}$, the shock evolution can be divided into three distinct phases characterized by different scaling relations:

\textit{(1) Free expansion phase.} Initially, the swept-up mass is much smaller than the ejecta mass, i.e., $M_{\rm s}\ll M_{\rm ej}$. The shock dynamics is dominated by the initial ejecta motion, and the shock velocity remains nearly constant: $v_{\rm s}\propto t^{0}.$

\textit{(2) Interaction phase.} 
When the swept-up mass becomes non-negligible, the interaction between ejecta and ambient gas significantly affects the shock evolution, while the ejecta still dominates the overall dynamics ($M_{\rm s}<M_{\rm ej}$). A double-shock structure is established, consisting of a forward shock propagating into the surrounding medium and a reverse shock propagating back into the ejecta. In this work, we only consider the forward shock. The self-similar evolution of this phase has been extensively studied~\citep{chevalier1982self,Nadyozhin1985,chevalier1994emission,chevalier2011shock}. For ejecta and ambient density profiles characterized by indices $n$ and $s$, respectively, the shock velocity evolves as $v_{\rm s}\propto t^{-\frac{3-s}{n-s}}$ with time or $v_{\rm s}\propto R^{-\frac{3-s}{n-3}}_{\rm s}$ with shock radius.

\textit{(3) Sedov-Taylor phase.} 
When the swept-up mass becomes comparable to the ejecta mass, the reverse shock has propagated through the entire ejecta, and the explosion enters the energy-conserving Sedov--Taylor (ST) phase. In this stage, the dynamics are determined mainly by the swept-up ambient medium. The shock velocity can be derived from energy conservation and follows $v_{\rm s}\propto t^{-\frac{3-s}{5-s}}$ with time or $v_{\rm s}\propto R^{-\frac{3-s}{2}}_{\rm s}$ with shock radius~\citep{Sedov1946,geoffrey1950formation}.

For an AGN disk, although the vertical gas density follows a Gaussian profile, the supernova ejecta are initially launched near the disk midplane, where the ambient density can be treated as $\rho_{\rm gas}\propto R^{0}$, i.e., $\rho_{\rm d}(z)\simeq \rho(r)$, and the shock evolution can be described by the self-similar scalings discussed above. For the detailed description of SN ejecta evolution in a uniform ambient medium, 
\cite{truelove1999evolution} developed an analytical solution for the shock 
velocity evolution during the ejecta-dominated (ED) stage, which provides a 
smooth transition from the initial free expansion phase to the subsequent 
interaction phase. The (forward) shock velocity in the ED stage is expressed as
\begin{equation}
\label{eq:ed}
\begin{aligned}
        v^*_{\rm ED}&=(\frac{2}{\alpha})^{\frac{1}{2}}\\
        &\times l_{\rm ED}(\frac{\{1-[(3-n)/3](\phi_{\rm ED}/l_{\rm ED}f_{\rm n})^{1/2}R^{*3/2}_{\rm s}\}^{\frac{5-n}{3-n}}}{1+(n/3)(\phi_{\rm ED}l_{\rm ED}f_{\rm n})^{1/2}R^{*3/2}_{\rm s}}),
\end{aligned}   
\end{equation}
where $R_{\rm s}^{*}=R_{\rm s}/R_{\rm ch}$ is the dimensionless shock radius. 
The characteristic radius and time scales are defined as 
$R_{\rm ch}=3.07(M_{\rm ej}/M_{\odot})^{1/3}n_{\rm gas}^{-1/3}\rm~pc$ and 
$t_{\rm ch}=423(E_{\rm ej}/10^{51}\rm~erg)^{-1/2}
(M_{\rm ej}/M_{\odot})^{5/6}n_{\rm gas}^{-1/3}\rm~yr$, respectively, 
where $n_{\rm gas}$ is the number density of the ambient medium. 
The corresponding dimensionless time is given by $t^{*}=t/t_{\rm ch}$. The parameters $\alpha$ and $f_{\rm n}$ depend on the ejecta properties and 
are defined as $\alpha=E_{\rm ej}/(M_{\rm ej}v_{\rm max}^{2}/2)$ and 
$f_{\rm n}=\frac{3}{4\pi}
\left[\frac{1-n/3}{1-(n/3)(v_{\rm t}/v_{\rm max})^{3-n}}\right]$, 
respectively. The parameters $\phi_{\rm ED}$ and $l_{\rm ED}$ are determined 
by the outer ejecta density index $n$ and can be obtained through interpolation 
of the tabulated values listed in Table~\ref{tab:star}.
\begin{figure}[t]
\centering
\includegraphics[width=1.0\linewidth,height=0.8\linewidth]{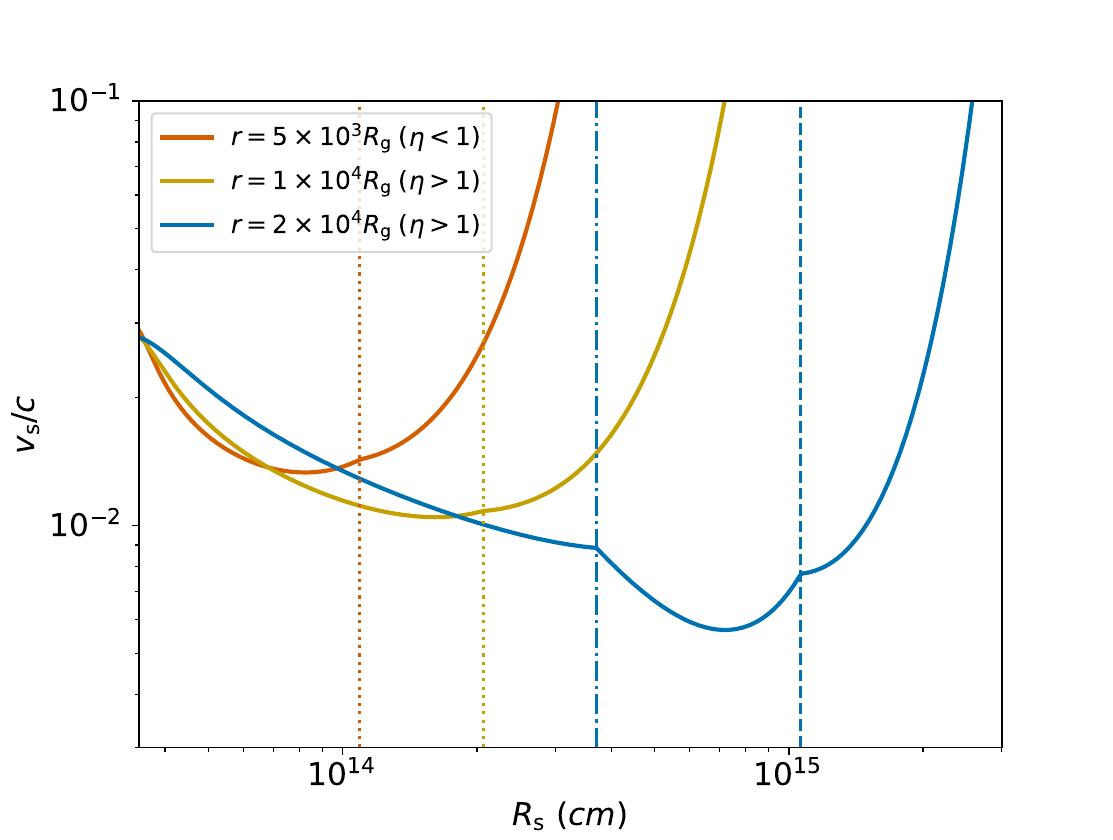}
\caption{Shock velocity as a function of shock radius $R_{\rm s}$ for $M_{\rm BH}=10^7M_{\odot}$, $E_{\rm ej}=10^{51}~{\rm erg}$, $M_{\rm ej}=10M_{\odot}$, $R_0=500R_{\odot}$, and $n=12$. Different colors correspond to different SN explosion radii in the disk. The dotted, dash-dotted, and dashed vertical lines indicate $R_{\rm ED}$, $R_{\rm tr}$, and $R_{\rm ST}$, respectively.}
\label{fig:v}
\end{figure}
We define the transition radius $R_{\rm tr}$ as the radius at which the swept-up 
mass equals the ejecta mass, i.e., $M_{\rm s}=M_{\rm ej}$. Beyond this point, the 
shock evolution enters the ST phase, where the shock velocity 
follows the scaling relation $v_{\rm s}\propto R_{\rm s}^{-3/2}$. Therefore, the 
shock velocity throughout the entire evolution can be approximated by the 
following piecewise function:

\begin{equation}
\label{eq:uniform}
v_{\rm s}=
\begin{cases}
v_{\rm ED}(R_{\rm s}), & R_{\rm s}<R_{\rm tr},\\
v_{\rm ED}(R_{\rm s})\left(\frac{R_{\rm s}}{R_{\rm tr}}\right)^{-3/2},
& R_{\rm s}\geq R_{\rm tr}.
\end{cases}
\end{equation}

We note that for an SN occurring at a radial position $r$ in the disk, the 
available gas mass at this location is limited and may be smaller than the 
ejecta mass $M_{\rm ej}$. In this case, the shock may never enter the 
ST phase. The total swept-up gas mass is given by $M_{\rm tot}(r)=\int_{R_0}^{\infty}\rho_{\rm d}(z)4\pi z^2 dz,$
where $R_0$ is the initial shock radius, which is set to the progenitor radius. 
We define the ratio $\eta=M_{\rm tot}/M_{\rm ej}$. If $\eta>1$, the swept-up 
mass is sufficient for the shock to enter the ST phase, and Eq (\ref{eq:uniform}) 
can be applied. Otherwise, the ejecta-dominated solution given by 
Eq (\ref{eq:ed}) describes the entire shock evolution.

However, the disk density is not real uniform in the vertical direction, especially when the shock approaches the disk surface. The decreasing density gradient can lead to an additional acceleration of the shock. Following the self-similar solution for shock propagation in steep density gradients~\citep{Sakurai1960}, the shock velocity scales as $v_{\rm s}\propto \rho_{\rm gas}^{-\mu}$, where $\mu\simeq0.19$ provides a good approximation to numerical simulations of shock breakout from stellar envelopes~\citep{matzner1999expulsion}. To incorporate this acceleration effect, we introduce a correction factor $\exp(\mu' R_{\rm s}^2/2H^2)$ into the time evolution of the shock radius in Eq.~(\ref{eq:ed}) and Eq.~(\ref{eq:uniform}), where $\mu'=\mu+1/n$ for the ED stage and $\mu'=\mu+1/5$ for the ST stage. The coefficients $1/n$ and $1/5$ come from the scaling of $R_{\rm s}\propto \rho^{1/n}$ and $R_{\rm s}\propto \rho^{1/5}$ for interaction phase and ST phase, respectively. This correction compensates for the enhanced deceleration caused by the increasing swept-up mass and reduces the decline of the shock velocity. At sufficiently large heights, the acceleration induced by the density gradient can eventually dominate over the deceleration, causing the shock to enter an acceleration phase. Following the analysis of \citet{grishin2021supernova}, this transition occurs at the minimum shock velocity. For instance, if the transition takes place during the ST phase, where the modified velocity evolution is $v_{\rm s}\propto R_{\rm s}^{-3/2}\exp(\mu R_{\rm s}^{2}/2H^2)$, the transition radius can be obtained by setting $\mathrm{d}\ln v_{\rm s}/\mathrm{d}\ln R_{\rm s}=0$, yielding $R_{\rm ST}=\sqrt{3H^2/(2\mu)}$. Similarly, for the ED phase, where $v_{\rm s}\propto R_{\rm s}^{-3/(n-3)}\exp(\mu R_{\rm s}^{2}/2H^2)$, the corresponding transition radius is $R_{\rm ED}=\sqrt{3H^2/[(n-3)\mu]}$.

In summary, for $\eta \lesssim 1$, the swept-up mass is insufficient to trigger the ST phase, and the shock remains in the ejecta-dominated ED regime throughout its evolution. The shock velocity is therefore expressed as
\begin{equation}
\label{eq:eta_less}
v_{\rm s}=
\begin{cases}
v_{\rm ED}(R_{\rm s})f_{\rho}(R_{\rm s}),
& R_{\rm s}<R_{\rm ED},\\[6pt]
v_{\rm ED}(R_{\rm ED})f_{\rho}(R_{\rm ED})
\left[\frac{\rho_{\rm d}(R_{\rm s})}
{\rho_{\rm d}(R_{\rm ED})}\right]^{-\mu},
& R_{\rm s}\geq R_{\rm ED}.
\end{cases}
\end{equation}
We define the density correction factor as $f_{\rho}(R)=
\left[\frac{\rho_{\rm d}(R)}{\rho}\right]^{-(\mu+1/n)}$. For $\eta>1$, the swept-up mass exceeds the ejecta mass, and the shock 
eventually enters the ST phase.\footnote{There is another case in which $\eta>1$ but $R_{\rm tr}>R_{\rm ED}$. In this case, the shock enters the acceleration phase directly from the ED phase before the swept-up mass reaches $M_{\rm ej}$, even though the total available mass in the disk is sufficient to satisfy $\eta>1$ (as in the case of $r=10^4R_{\rm g}$ in Fig.~\ref{fig:v}). The shock velocity therefore follows the same equations as those given in Eq.~(\ref{eq:eta_less}).}

The shock velocity evolution can be described by

\begin{equation}
\label{eq:eta_greater}
v_{\rm s}=
\begin{cases}
v_{\rm ED}(R_{\rm s})f_\rho(R_{\rm s}),
& R_{\rm s}<R_{\rm tr},\\[6pt]

A_{\rm tr}
\left(\frac{R_{\rm s}}{R_{\rm tr}}\right)^{-\frac{3}{2}}
\left[\frac{\rho_{\rm d}(R_{\rm s})}
{\rho_{\rm d}(R_{\rm tr})}\right]^{-(\mu+\frac{1}{5})},
& R_{\rm tr}\leq R_{\rm s}<R_{\rm ST},\\[6pt]

A_{\rm tr}
\left(\frac{R_{\rm s}}{R_{\rm tr}}\right)^{-\frac{3}{2}}
\left[\frac{\rho_{\rm d}(R_{\rm s})}
{\rho_{\rm d}(R_{\rm tr})}\right]^{-(\mu+\frac{1}{5})}
\left[\frac{\rho_{\rm d}(R_{\rm s})}
{\rho_{\rm d}(R_{\rm ST})}\right]^{-\mu},
& R_{\rm s}\geq R_{\rm ST}.
\end{cases}
\end{equation}
where we define the normalization factor at the transition radius as $A_{\rm tr}=v_{\rm ED}(R_{\rm tr})f_{\rho}(R_{\rm tr}).$

In Fig.~\ref{fig:v}, we show the shock velocity evolution for different SN explosion radii in an AGN disk around a $10^7M_{\odot}$ black hole, adopting typical parameters for a red supergiant explosion, $E_{\rm ej}=10^{51}~{\rm erg}$, $M_{\rm ej}=10M_{\odot}$, $R_0=500R_{\odot}$, and $n=12$. For an explosion at $r=5\times10^3R_{\rm g}$, the total mass available in the disk is smaller than the ejecta mass ($\eta<1$). The shock therefore transitions directly from the ED phase to the acceleration phase at $R_{\rm ED}$. In contrast, for an explosion at $r=2\times10^4R_{\rm g}$, the swept-up mass is sufficient to exceed the ejecta mass ($\eta>1$), allowing the shock to enter the ST phase at $R_{\rm tr}$ and subsequently transition to the acceleration phase at $R_{\rm ST}$.

\subsection{Light Curve}
\begin{figure*}[htbp]
\centering
\includegraphics[width=1.0\linewidth,height=0.35\linewidth]{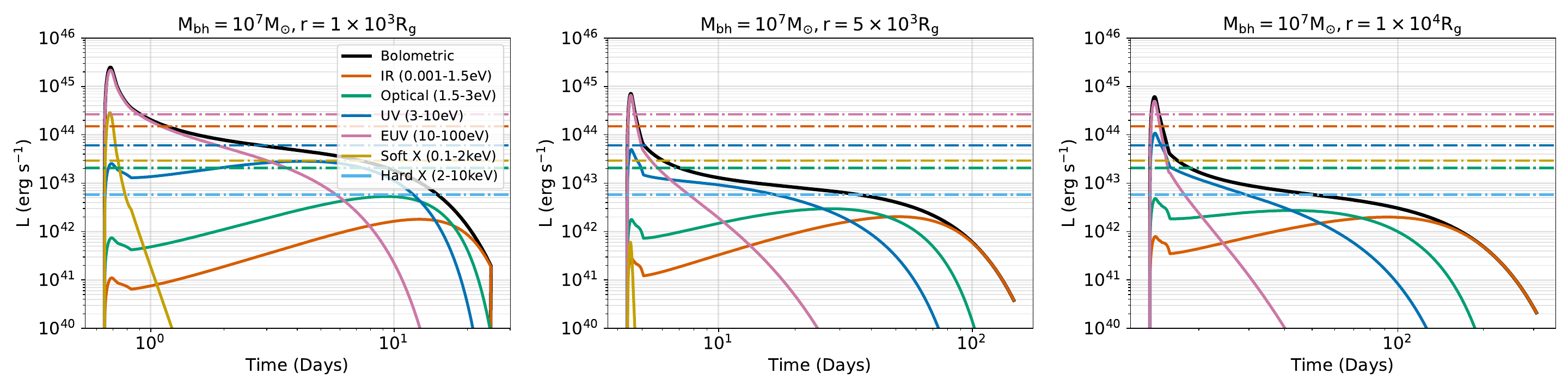}
\includegraphics[width=1.0\linewidth,height=0.35\linewidth]{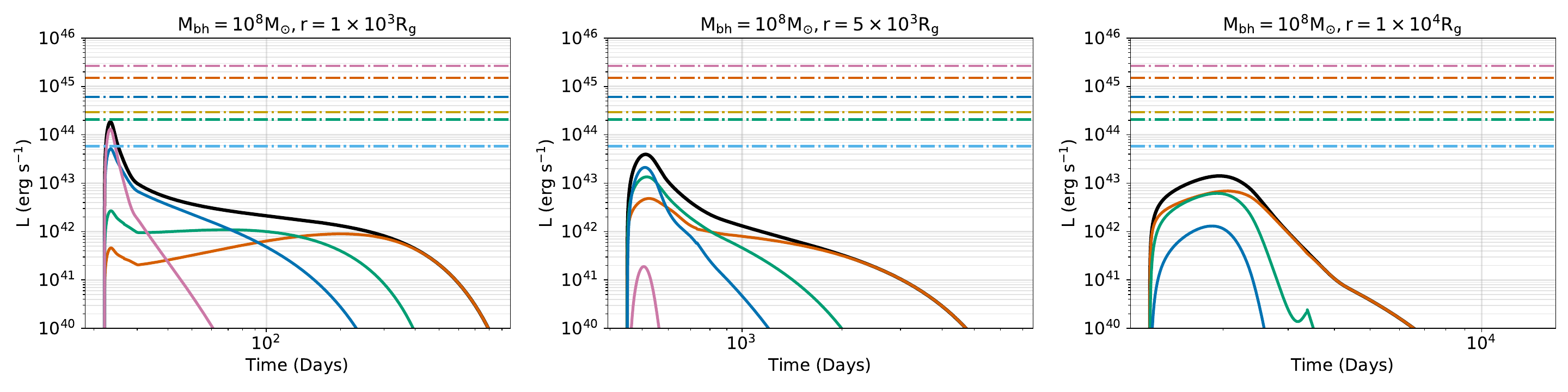}
\caption{Bolometric and multi-band light curves for $E_{\rm ej}=10^{51}~{\rm erg}$, $M_{\rm ej}=10M_{\odot}$, $R_0=500R_{\odot}$, and $n=12$, with different panels showing different SN explosion radii and black hole masses. The AGN background emission in each band is indicated by a horizontal dotted line.}
\label{fig:lc}
\end{figure*}
At the beginning of the shock evolution, the shock front is deeply embedded in the disk, where the large optical depth causes the photon diffusion velocity to be smaller than the shock velocity. Consequently, electromagnetic signals cannot escape ahead of the shock until it reaches the breakout height, $R_{\rm bo}$. This height can be estimated by requiring the optical depth ahead of the shock to satisfy~\citep{nakar2010early,svirski2012optical,margalit2022optical}
\begin{equation}
\tau_{\rm d}(R_{\rm bo})
= \int_{R_{\rm bo}}^{\infty} \kappa \rho_{\rm d}(z),{\rm d}z
= \frac{c}{v_{\rm s}}.
\end{equation}
When the shock reaches $R_{\rm bo}$, photons trapped behind the shock begin to escape, producing the shock breakout emission. Subsequently, the shocked gas cools and gives rise to the cooling emission. Meanwhile, the shock continues to propagate outward and interact with the surrounding matter, powering sustained interaction emission. The different phases and characteristic emission components of SNe interacting with a circumstellar medium are summarized in \cite{khatami2024landscape}. In general, photons originate from the region immediately behind the shock and from the downstream cooling gas. As they diffuse ahead of the shock, these photons interact with electrons in the unshocked material and can reprocess the radiation, thereby regulating the radiation temperature, primarily through free-free emission and absorption~\citep{nakar2010early,svirski2012optical}. To capture both the light-curve shape and the peak photon energy, we develop a one-zone model that describes the evolution of the radiation energy, photon number, and electron energy in the unshocked region,
\begin{equation}
\label{eq:lc}
\begin{aligned}
    \frac{dE_{\mathrm{rad}}}{dt}&=-\frac{E_{\mathrm{rad}}}{t_{\mathrm{diff}}}+L_{\mathrm{s}}+L_{\mathrm{sg}}+Q_{\mathrm{com}}+Q_{\mathrm{ff}}+6\pi R_{\mathrm{s}}^{2}v_{\mathrm{s}}n_{\mathrm{e}}kT_{\mathrm{e}},\\
    \frac{dN_{\mathrm{rad}}}{dt}&=-\frac{N_{\mathrm{rad}}}{t_{\mathrm{diff}}}+\frac{L_{\mathrm{s}}}{3kT_{s}}+\frac{L_{\mathrm{sg}}}{3kT_{sg}}+\dot{N}_{\mathrm{ff}},\\
    \frac{dE_{\mathrm{e}}}{dt}&=-Q_{\mathrm{com}}-Q_{\mathrm{ff}}-6\pi R_{\mathrm{s}}^{2}v_{\mathrm{s}}n_{\mathrm{e}}kT_{\mathrm{e}},\\
    Q_{\mathrm{com}}&=4\sigma _{\mathrm{T}}\bar{n}_{\mathrm{e}}\frac{k\left( T_{\mathrm{e}}-T_{\mathrm{rad}} \right)}{m_{\mathrm{e}}c}E_{\mathrm{rad}},\\
    Q_{\mathrm{ff}}&=c\kappa\bar{\rho_{e}}  _{\mathrm{ff}}\left( aT_{\mathrm{e}}^{4}V-E_{\mathrm{rad}} \right)\\
    \dot{N}_{\mathrm{ff}}&=c\kappa\bar{\rho_{e}} _{\mathrm{ff}}\left( \frac{aT_{\mathrm{e}}^{4}V}{3kT_{\mathrm{e}}}-N_{\mathrm{rad}} \right),
\end{aligned}
\end{equation}
where the photon diffusion timescale in the unshocked region is $ t_{\rm diff}=\int_{R_{\rm s}}^{\infty}(\kappa\rho_{\rm d}/c)(z-R_{\rm s}){\rm d}z$. The photons injected into the unshocked region originate from two sources: the shock and the downstream cooling gas. The luminosity injected by the shock is $L_{\rm s}=2\pi\rho_{\rm d}v_{\rm s}^{3}R_{\rm s}^{2}$, with a characteristic photon temperature of $T_{\rm s}=(18\rho_{\rm d}v_{\rm s}^{2}/7a)^{1/4}$~\citep{zhang2018physics,chen2025observational}, corresponding to the typical post-shock temperature of a radiation-dominated shock. For the cooling gas, we calculate the injected luminosity $L_{\rm sg}$ and the corresponding photon temperature $T_{\rm sg}$ by adopting a shell geometry, uniform expansion, and a blackbody radiation spectrum for the shocked gas; the detailed treatment follows Appendix A of \cite{margalit2022analytic}. Before the shock reaches the photosphere $R_{\rm ph}$, defined by $\tau_{\rm d}=1$, the diffusing photons interact with electrons in the unshocked region. We take the volume of this region to be $V=4\pi(R_{\rm ph}^{3}-R_{\rm s}^{3})/3$. Initially, the electrons are assumed to be in thermal equilibrium with the protons and therefore have a temperature $T_{\rm e}=T(r)$ given by the disk structure model. The term $Q_{\rm com}$ describes the energy exchange between the radiation field and electrons through Compton scattering, where $\bar{n}_{\rm e}$ is the average electron number density in the unshocked region. The terms $Q_{\rm ff}$ and $\dot{N}_{\rm ff}$ describe the changes in the radiation energy and photon number, respectively, due to free-free emission and absorption, with $\kappa_{\rm ff}\simeq2\times10^{24}\bar{\rho}_{\rm e}T_{\rm e}^{-3.5}~{\rm cm}^2~{\rm g}^{-1}$~\citep{rybicki1979radiative,mckinney2017double}. The last term in the radiation-energy equation accounts for the energy transferred from the shock-heated electrons to the radiation field through adiabatic expansion of the shocked gas.

We solve Eq.~(\ref{eq:lc}) starting from the breakout time $t_{\rm bo}$. The observed bolometric luminosity is then given by $L_{\rm bol}=E_{\rm rad}/t_{\rm diff}$, while the characteristic radiation temperature is $T_{\rm rad}=E_{\rm rad}/(3kN_{\rm rad})$. Here, we assume that the emergent spectrum has a blackbody shape with a characteristic temperature $T_{\rm rad}$. Before the shock reaches $R_{\rm ph}$, efficient free-free emission can produce a sufficient number of photons to maintain approximate thermal equilibrium, such that the resulting $T_{\rm rad}$ naturally approaches the corresponding blackbody temperature. If photon production is insufficient, however, the radiation field becomes photon-starved and the resulting spectrum is harder than a blackbody. Once the shock reaches the photosphere $R_{\rm ph}$, the emitting region itself becomes directly visible. At this stage, Eq.~(\ref{eq:lc}) naturally accounts for the luminosity and characteristic temperature of the directly observed emission sources.

In Fig.~\ref{fig:lc}, we present an example of the bolometric and corresponding multi-band light curves for a supernova occurring in an accretion disk around a $10^7M_{\odot}$ black hole, adopting typical parameters for an RSG progenitor. Equation~(\ref{eq:lc}) captures the main features expected for a supernova light curve, namely a rapid initial rise followed by an extended plateau phase. Despite the simplified treatment of the radiation transport, our results are broadly consistent with those of previous studies of SNe in AGN disks, such as \citet{grishin2021supernova}, which employ hydrodynamic simulations with \textit{SNEC} or \textit{HORMONE}~\citep{morozova2015light,hirai2016hyperbolic}. In particular, the shock breakout time, peak luminosity, and cooling duration obtained from our model are in reasonable agreement with their results, supporting the validity of our semi-analytic treatment. As a general trend across different combinations of $r$ and $M_{\rm BH}$, a larger black hole mass or a more distant explosion location tends to produce a lower peak luminosity, a longer light-curve duration, and a lower characteristic photon energy. This trend is qualitatively consistent with previous studies of SNe embedded in disks~\citep{li2023core}.

We also show the representative AGN background emission in different bands, obtained by integrating the background SED over the corresponding energy ranges. We adopt the empirical prescription of \citet{ho2008nuclear} to model the SED of AGN, whose shape is determined by the Eddington ratio, $l_{\rm E}=L_{\rm bol}/L_{\rm Edd}$. Since $l_{\rm E}$ is also an input parameter of our SG disk model, this treatment provides a self-consistent estimate of the AGN background for each disk configuration. The resulting SED is normalized to the bolometric luminosity, $L_{\rm bol}$. As shown in Fig.~\ref{fig:lc}, for most explosion locations and black hole masses, the SN emission in most bands is outshone by the underlying AGN background for the majority of the evolution. Nevertheless, in some cases, the SN emission can exceed the AGN background in specific bands around the bolometric peak, making the transient potentially detectable. This suggests that SNe occurring in AGN disks may be preferentially searched for in specific bands and around AGNs with particular black hole masses (also $l_{\rm E}$, which affects both disk and background radiation). We explore this band- and AGN-dependent detectability in detail in the next section.

\begin{table}[t]
\centering
\renewcommand{\arraystretch}{1.2}
\setlength{\tabcolsep}{10pt}  
\caption{The adopted fiducial progenitor properties, including the
explosion energy $E_{\rm ej}$, ejecta mass $M_{\rm ej}$, stellar
radius $R_0$, and outer ejecta density index $n$. The adopted values
are representative of RSG, YSG, BSG, and WR progenitors and are
motivated by previous studies of shock-cooling and stripped-envelope
supernovae~\citep{ganot2016detection,tanaka2009type,bersten2012type}.}
\label{tab:star}
\begin{tabular}{lcccc}
\toprule
Progenitors & RSGs & YSGs & BSGs & WRs  \\
\midrule
$E_{\rm ej}$~(erg) &$10^{51}$ &$10^{51}$ &$10^{51}$  &$10^{51}$  \\
$M_{\rm ej}$~($M_{\odot}$) &10 &3 &20 &3  \\
$R_0$~($R_{\odot}$) &500 &200 &40 & 1\\
$n$ &12 &10 &10 &10\\
\bottomrule
\end{tabular}
\end{table}

\section{Results}

\subsection{Observability of SNe in AGN Disks}
\label{sec:obs}
\begin{figure*}[htbp]
\centering
\includegraphics[width=1.0\linewidth,height=0.35\linewidth]{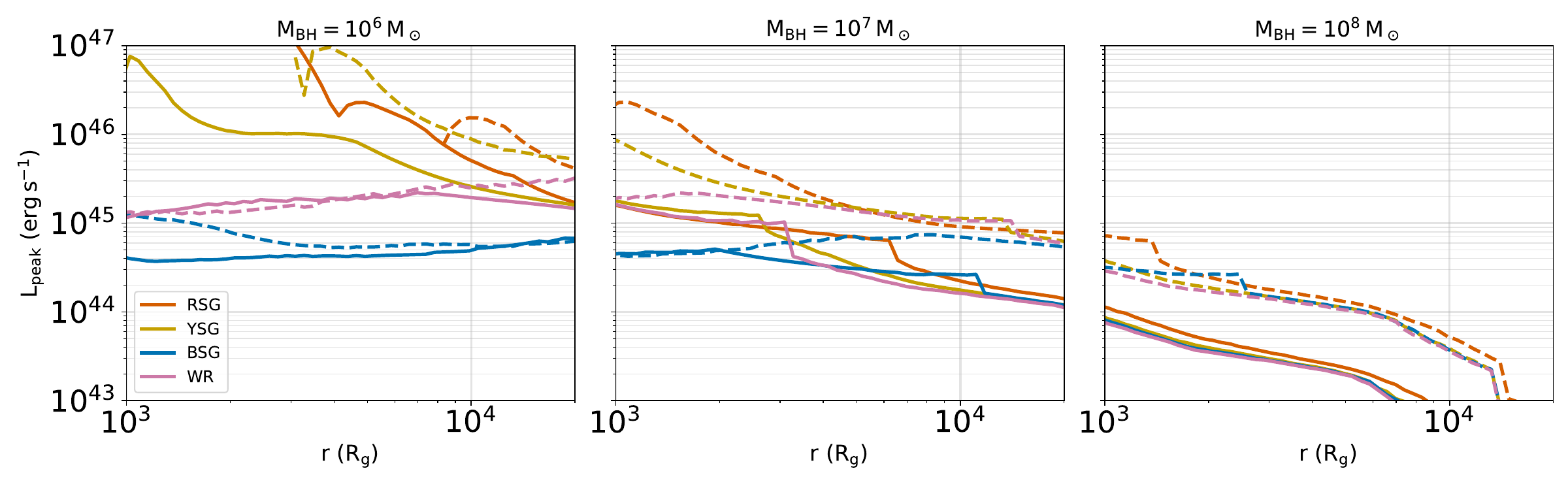}
\includegraphics[width=1.0\linewidth,height=0.35\linewidth]{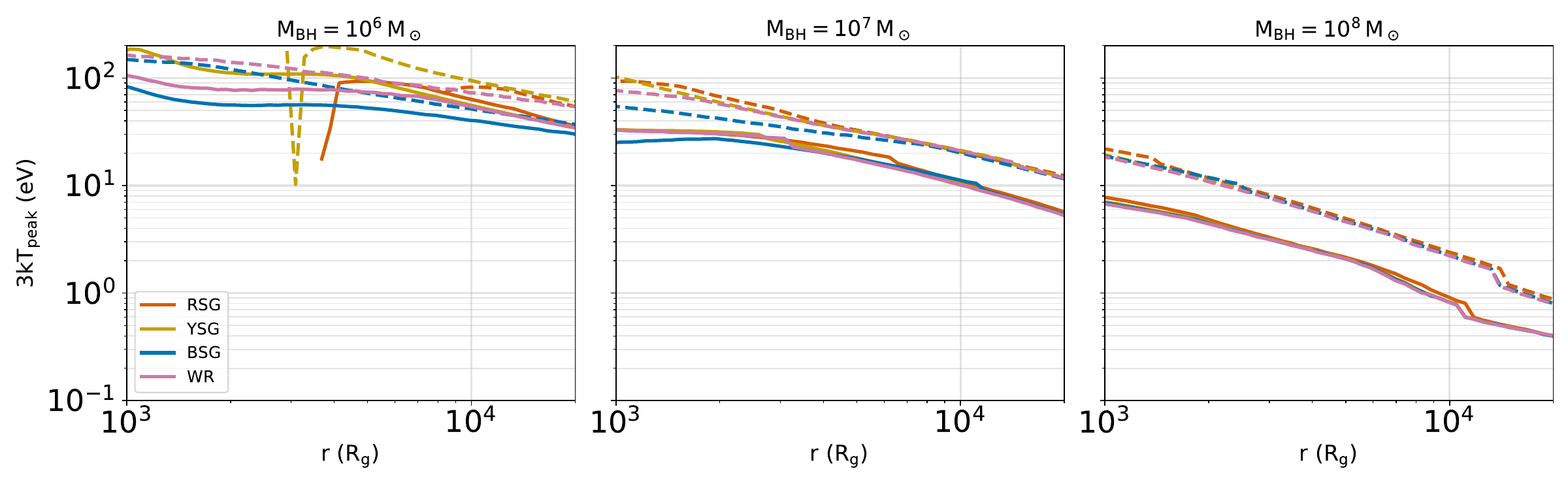}
\caption{Peak bolometric luminosity and the corresponding characteristic photon energy as a function of the explosion radius in the disk for different types of SN progenitors. The solid and dashed lines correspond to $l_{\rm E}=1$ and $l_{\rm E}=0.1$, respectively.}
\label{fig:lt}
\end{figure*}

From the example light curves presented above, we find that the period around the peak luminosity, $L_{\rm peak}$, provides a particularly important observational window for SNe occurring in AGN disks. We therefore investigate the peak luminosity and the corresponding characteristic temperature, $T_{\rm peak}$, for SNe exploding at radii of $r=10^3$--$10^4R_{\rm g}$, which broadly cover the region where stars can form or become embedded in AGN accretion disks~\citep{goodman2003self,sirko2003spectral,jermyn2022effects,fan2023situ}. The results are presented in Fig.~\ref{fig:lt}. Given the uncertainties in the properties of stellar progenitors embedded in AGN disks, we consider four representative progenitor types: red supergiants (RSGs), yellow supergiants (YSGs), blue supergiants (BSGs), and Wolf--Rayet (WR) stars. The adopted progenitor properties and explosion energies are summarized in Table~\ref{tab:star}. 

From Fig.~\ref{fig:lt}, we find that for a relatively low-mass black hole, $M_{\rm BH}=10^6M_{\odot}$, RSGs and YSGs produce substantially higher $L_{\rm peak}$ than BSGs and WRs at $r<10^4R_{\rm g}$. This progenitor dependence becomes progressively weaker with increasing $M_{\rm BH}$, as the more extended disk around a more massive black hole increasingly washes out the intrinsic properties of the progenitor. The dependence of $T_{\rm peak}$ on progenitor type is much weaker, even at $M_{\rm BH}=10^6M_{\odot}$, and becomes less pronounced toward larger $M_{\rm BH}$. Thus, SNe occurring in disks around lower-mass black holes may retain more prominent signatures of their progenitor properties, potentially allowing the progenitor type to be better constrained from observations.

We also compare the results for $l_{\rm E}=0.1$ and $l_{\rm E}=1$ in Fig.~\ref{fig:lt}. The $l_{\rm E}=0.1$ models generally exhibit both higher $L_{\rm peak}$ and higher $T_{\rm peak}$. At the same time, the underlying AGN emission is weaker at lower  $l_{\rm E}$, reducing the contrast between the SN and the AGN background in the observed light curves. These trends suggest that low-luminosity AGNs, particularly those hosting relatively low-mass BHs, may provide promising environments in which to search for and characterize SNe occurring in AGN accretion disks.

\begin{figure*}[htbp]
\centering
\includegraphics[width=0.9\linewidth,height=0.8\linewidth]{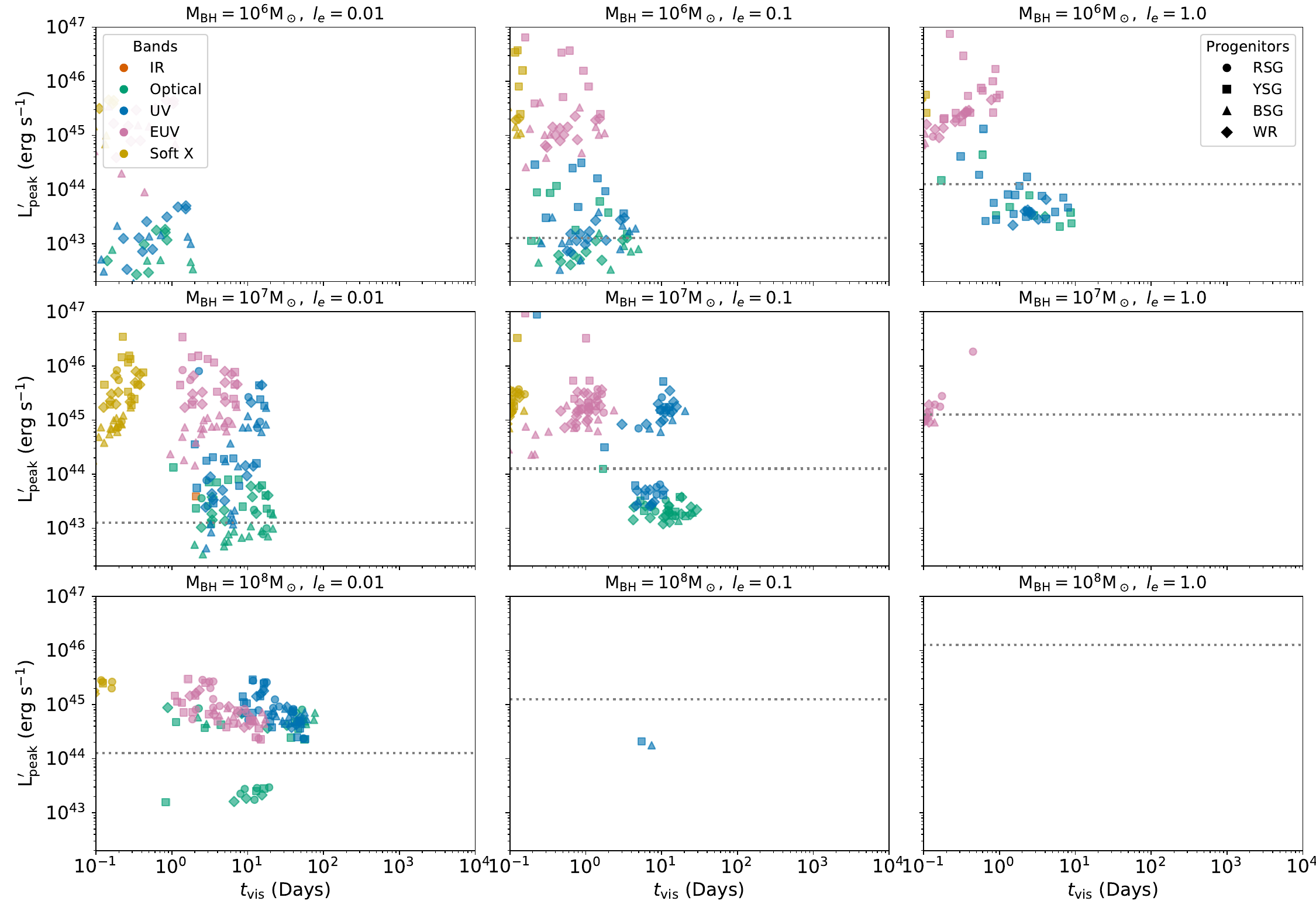}
\caption{Distributions of SNe in the visibility timescale--peak luminosity plane for different combinations of BH mass \(M_{\rm BH}\) and Eddington ratio \(l_{\rm E}\). Different marker shapes denote different progenitor types, while different colors denote different electromagnetic bands. Here, \(L^{'}_{\rm peak}\) is the peak luminosity in the corresponding band, and \(t_{\rm vis}\) is the timescale over which the SN emission exceeds the AGN background emission in that band. The AGN bolometric luminosities are indicated by dotted lines for reference.}

\label{fig:sct}
\end{figure*}

As discussed above, the observability of SNe occurring in AGN disks depends on the black hole mass, AGN Eddington ratio, and progenitor properties. To investigate the combined effects of these parameters, we perform a Monte Carlo parameter scan over the AGN and progenitor parameter space. We consider three black hole masses, \(M_{\rm BH}=10^6\), \(10^7\), and \(10^8\,M_\odot\), and three Eddington ratios, \(l_{\rm E}=0.01\), \(0.1\), and \(1\). For each combination of \((M_{\rm BH},l_{\rm E})\), we generate 50 realizations. In each realization, the SN progenitor is selected with equal probability from four representative classes: RSG, YSG, BSG, and WR. For the selected progenitor class, the ejecta mass \(M_{\rm ej}\), stellar radius \(R_0\), density-profile index \(n\), and explosion energy \(E_{\rm ej}\) are perturbed around their respective fiducial values. Specifically, \(M_{\rm ej}\), \(R_0\), and \(E_{\rm ej}\) are sampled from log-normal distributions centered on their fiducial values, with standard deviations of \(0.20\), \(0.30\), and \(0.15\) dex, respectively, while \(n\) is sampled from a normal distribution centered on its fiducial value with a standard deviation of \(0.5\). Independently, the explosion radius is sampled from a log-uniform distribution over \(10^3\)--\(10^{4}R_{\rm g}\). This procedure allows us to assess the robustness of the predicted observables against uncertainties in progenitor properties and explosion locations, rather than relying on a single set of fiducial parameters. For each realization, we calculate the peak luminosity, \(L^{\prime}_{\rm peak}\), and the visibility timescale, \(t_{\rm vis}\), in each of the considered electromagnetic bands. Here, \(L^{\prime}_{\rm peak}\) denotes the peak luminosity in a specific band and is therefore distinct from the bolometric peak luminosity \(L_{\rm peak}\) defined above. The visibility timescale \(t_{\rm vis}\) is defined as the duration over which the SN luminosity in a given band exceeds the AGN background luminosity in the same band. Thus, each realization contributes a set of \((L^{\prime}_{\rm peak},t_{\rm vis})\) values corresponding to the different electromagnetic bands. The resulting distributions in the \(t_{\rm vis}\)--\(L^{\prime}_{\rm peak}\) plane are shown in Fig.~\ref{fig:sct}.

As shown in Fig.~\ref{fig:sct}, AGNs with lower $M_{\rm BH}$ and lower $l_{\rm E}$ generally host a larger number of detectable SNe and provide a broader range of observable electromagnetic bands, extending from the IR to the soft X-ray band. In contrast, for combinations of high $M_{\rm BH}$ and high $l_{\rm E}$, such as $M_{\rm BH}=10^8M_{\odot}$ and $l_{\rm E}=1$, only a small fraction of SNe remain detectable, with the observable emission largely confined to the IR band. This trend is consistent with the discussion above, indicating that AGNs with lower luminosities and lower BH masses are generally more favorable targets for searches for SNe occurring in AGN disks. More specifically, for the lowest BH mass, $M_{\rm BH}=10^6M_{\odot}$, the number of detectable SNe decreases only gradually as $l_{\rm E}$ increases. In contrast, for higher BH masses, such as $M_{\rm BH}=10^7M_{\odot}$ and $10^8M_{\odot}$, only a small number of SNe remain detectable at high $l_{\rm E}$. A similar trend is found when varying $M_{\rm BH}$ at the lowest $l_{\rm E}$. These results suggest that a low value of either $M_{\rm BH}$ or $l_{\rm E}$ can largely preserve a relatively large detectable SN population, such that increasing the other parameter does not substantially suppress the detectability. Therefore, AGNs with either a low BH mass or a low Eddington ratio may be promising targets for searches for SNe across multiple electromagnetic bands.

For the different progenitor types considered here, we do not find any clear clustering in the distributions, nor is any particular progenitor type strongly associated with a specific electromagnetic band. The distributions of the different progenitor types also exhibit broadly similar variations with $M_{\rm BH}$ and $l_{\rm E}$. These results suggest that, for a fixed explosion energy, the detectability of disk SNe is not strongly dependent on the progenitor type in our model. In other words, the progenitor type does not appear to be a dominant factor in determining the observational prospects of SNe occurring in AGN disks, compared with the effects of the AGN properties.

\begin{figure*}[htbp]
\centering
\includegraphics[width=1.05\linewidth,height=0.35\linewidth]{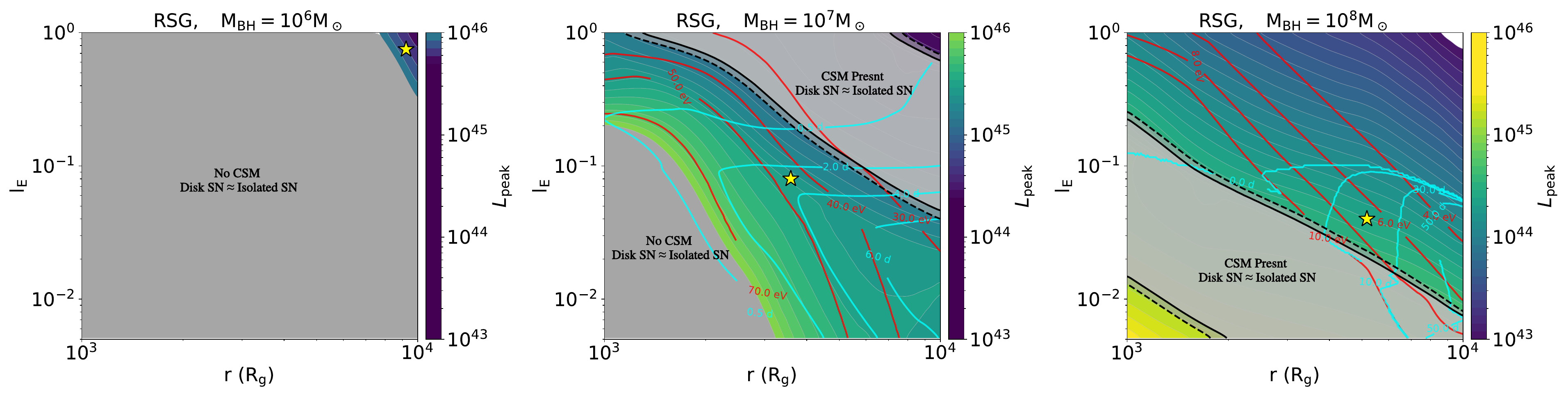}
\includegraphics[width=1.05\linewidth,height=0.35\linewidth]{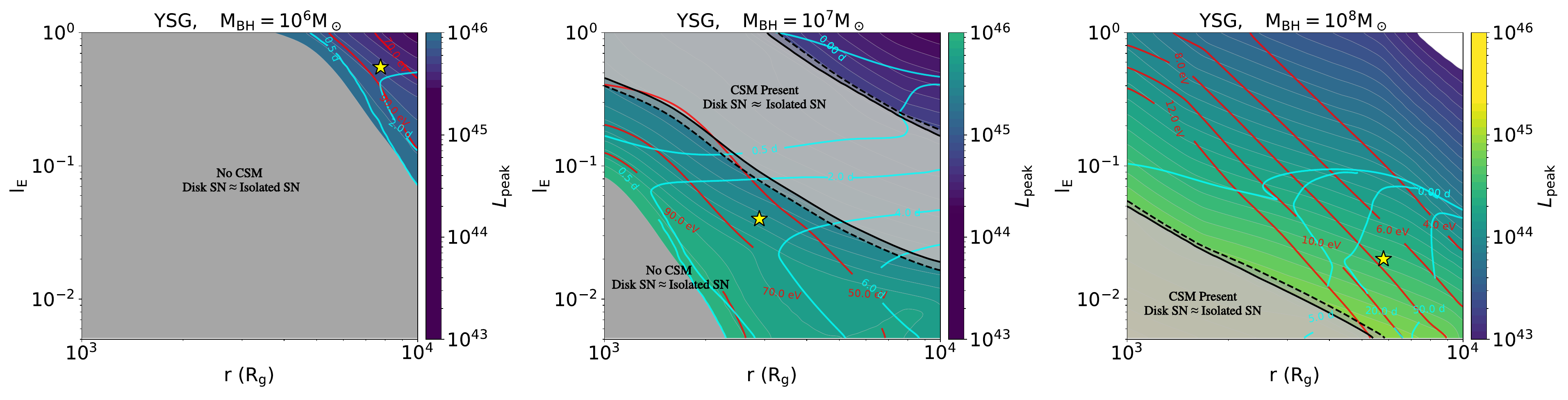}
\caption{
Contours of the peak luminosity $L_{\rm peak}$ of disk SNe in the $r$--$l_{\rm E}$ parameter space for RSG and YSG progenitors. The radiation temperature at the time of peak luminosity $T_{\rm peak}$ and visibility timescale $t_{\rm vis}$, defined as the duration over which the SN emission in the corresponding peak-temperature band exceeds the AGN background, are indicated by red and cyan contours, respectively. Gray-shaded regions bounded by solid and dashed lines indicate models indistinguishable from isolated CSM-SNe at the nominal 95\% and 99\% thresholds, respectively. Regions where disk SNe are indistinguishable from isolated SNe without CSM are also shaded gray and labeled accordingly. The yellow star marks a representative disk-SN model distinguishable from the isolated-SN reference grid.
}
\label{fig:ry}
\end{figure*}

\begin{figure*}[htbp]
\centering
\includegraphics[width=1.05\linewidth,height=0.35\linewidth]{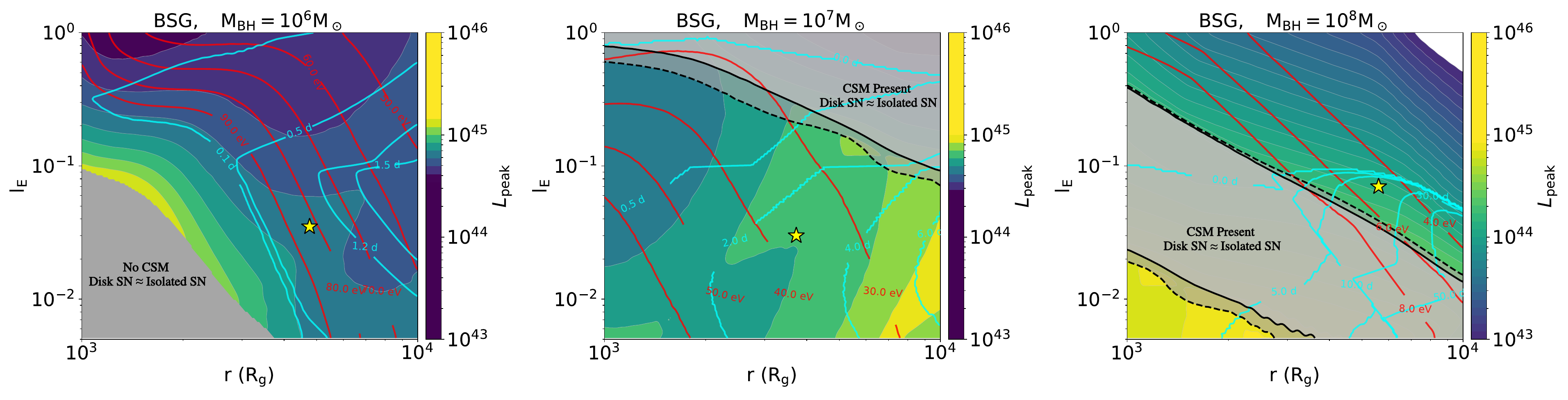}
\includegraphics[width=1.05\linewidth,height=0.35\linewidth]{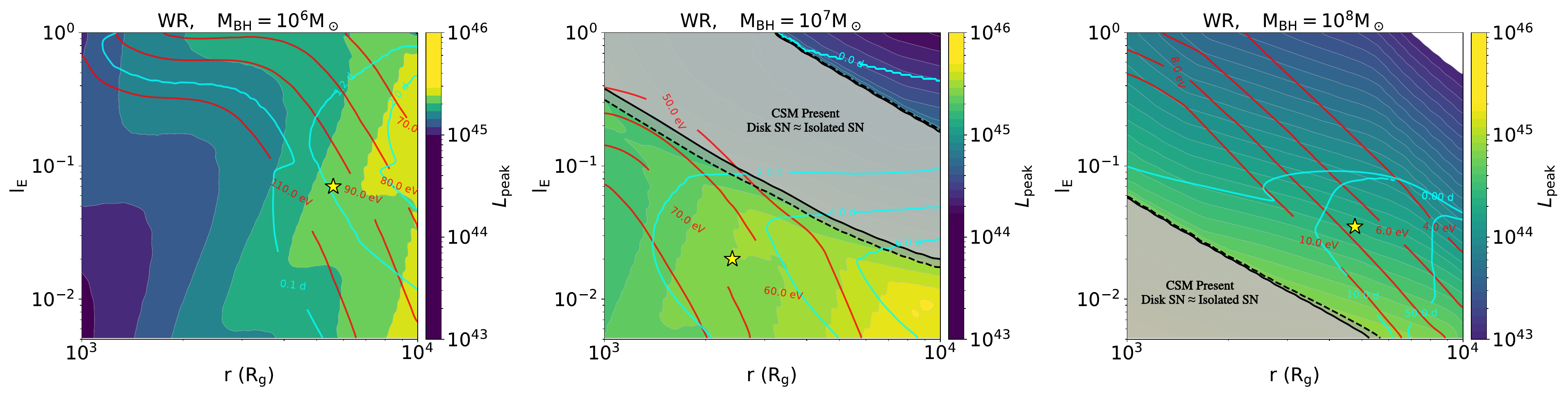}
\caption{Same as Fig.~\ref{fig:ry} but for BSG and WR progenitors.}
\label{fig:bw}
\end{figure*}

\subsection{Distinguishing AGN-disk SNe from Isolated SNe}
\label{sec:dis}
In Section~\ref{sec:obs}, we show that SNe occurring in AGN disks with lower $M_{\rm BH}$ or $l_{\rm E}$ are more likely to outshine the underlying AGN emission and thus become detectable. However, from a physical perspective, disk SNe are not fundamentally distinct from isolated SNe interacting with circumstellar material (CSM), as both are powered by the interaction between the SN ejecta and the surrounding matter. Therefore, distinguishing disk SNe from those occurring in the host galaxy outside the AGN disk remains an observational challenge. Rather than attempting to determine whether an individual SN is associated with an AGN disk, we instead seek observational signatures that are unique to disk SNe, such that their detection would provide strong evidence for a disk origin.
 The distinct signatures identified in this analysis may provide useful guidance for selecting and identifying candidate disk SNe in observations.

To investigate this issue systematically, we first construct a comprehensive reference database of isolated SNe interacting with CSM. This database is designed to encompass a broad range of physical CSM properties and to characterize the corresponding observable properties of isolated SNe, providing a systematic reference set against which the observational signatures of disk SNe can be compared. We calculate their light curves using the same method described in Section~\ref{sec:model}, ensuring a consistent treatment of the radiative processes in both environments. We adopt a broken power-law density profile for the CSM,
\begin{equation}
\rho_{\rm CSM}(R)\propto
\begin{cases}
R^{-2}, & R\leq R_{\rm CSM},\\
R^{-p}, & R>R_{\rm CSM},
\end{cases}
\end{equation}
where $R_{\rm CSM}$ is the characteristic radius at which the density profile steepens, and the density normalization is determined by the total CSM mass $M_{\rm CSM}$. We fix the ejecta outer density slope to $p=12$, and explore a broad parameter space spanning $R_{\rm CSM}=10^{14}\text{--}10^{19}\,\mathrm{cm}$ and $M_{\rm CSM}=10^{-5}\text{--}10^0\,M_{\rm ej}$. This range of CSM radii encompasses both compact circumstellar environments inferred from observations of Type II SNe~\citep{dessart2023using,bostroem2023early} and more extended circumstellar shells considered in theoretical studies. The adopted CSM mass range covers interaction regimes from negligible CSM effects to cases in which the CSM mass is comparable to the ejecta mass~\citep{singh2024unravelling,pan2013superluminous}.

We then perform a corresponding parameter scan for SNe exploding in AGN disks, covering $l_{\rm E}=10^{-3}\text{--}10^0$ and $r=10^3\text{--}10^4,R_{\rm g}$. We consider four progenitor types and compare the resulting disk-SN models against the isolated CSM-SN reference grid constructed above, in order to identify regions of the disk-SN parameter space that cannot be reproduced by isolated SNe interacting with CSM.

For this comparison, we use a three-dimensional observable space spanned by the peak luminosity $L_{\rm peak}$, the radiation temperature at the time of peak luminosity, $T_{\rm peak}$, and the characteristic decline timescale, $t_{\rm dec}$~\citep{arnett1979theory,arnett1980analytic,arnett1982type},
\begin{equation}
t_{\rm dec}=\frac{3\kappa M_{\rm bo}}{4\pi c v_{\rm bo}},
\end{equation}
which characterizes the duration of the light-curve plateau. Here, $M_{\rm bo}$ is the ejecta mass at breakout. To quantify the distinguishability of the two model populations, we calculate the distance between each disk-SN model and its nearest neighbor in the isolated CSM-SN grid. The distance is defined in logarithmic observable space as
\begin{equation}
\begin{aligned}
\chi^2_{\rm min}
&= \min_j \Bigg[
\left(\frac{\log L_{\rm peak}-\log L_{{\rm peak},j}^{\rm iso}}{\sigma_L}\right)^2\\
&\qquad+
\left(\frac{\log T_{\rm peak}-\log T_{{\rm peak},j}^{\rm iso}}{\sigma_T}\right)^2\\
&\qquad+
\left(\frac{\log t_{\rm dec}-\log t_{{\rm dec},j}^{\rm iso}}{\sigma_t}\right)^2
\Bigg].
\end{aligned}
\end{equation}
where $j$ indexes the isolated-SN models. We adopt $\sigma_L=\sigma_T=\sigma_t=0.1$ dex, corresponding to a characteristic tolerance of approximately $26\%$ in each observable. We use a $k$-d tree to efficiently identify the nearest isolated SN model to each disk-SN model in the three-dimensional observable space. The resulting minimum distance, $\chi^2_{\rm min}$, is then compared with the nominal 95th and 99th percentiles of a $\chi^2$ distribution with three degrees of freedom, corresponding to $\chi^2_{\rm min}=7.815$ and $11.345$, respectively. Models below these thresholds have isolated-SN counterparts within the adopted tolerance, while those with $\chi^2_{\rm min}>11.345$ are considered distinguishable from the isolated SN reference grid under our criterion. Because this statistic is based on nearest-neighbor distances to a finite model grid, the thresholds are used as measures of separation rather than formal confidence levels.

In some cases, the progenitor radius exceeds the disk scale height, i.e., $R_0>H$, and the surrounding disk mass is too small for the progenitor to be effectively embedded in the AGN disk. In these cases, the resulting emission is expected to be indistinguishable from that of an SN {\it without} CSM interaction. To account for this possibility, we classify disk-SN models with a surrounding disk mass below $10^{-5}M_{\rm ej}$ (the lower limit of our isolated CSM-SN mass grid) as directly indistinguishable from isolated SNe without CSM.

In Fig.~\ref{fig:ry} and Fig.~\ref{fig:bw}, we present contours of the peak luminosity $L_{\rm peak}$, peak temperature $T_{\rm peak}$, and visibility timescale $t_{\rm vis}$ of disk SNe in the $r$--$l_{\rm E}$ parameter space for the four progenitor types. Here, $t_{\rm vis}$ is defined as the duration over which the SN emission in the band corresponding to $T_{\rm peak}$ exceeds the underlying AGN background. For $M_{\rm BH}=10^6,M_\odot$, most disk SNe are observationally indistinguishable from isolated SNe with negligible CSM interaction. This is particularly the case for progenitors with larger initial radii $R_0$, such as RSGs and YSGs, which are only weakly embedded in the disk. In contrast, disk SNe from WR progenitors are more readily distinguished from isolated SNe. As $M_{\rm BH}$ increases, disk SNe become predominantly indistinguishable from isolated SNe with substantial CSM interaction. The indistinguishable regions in the $r$--$l_{\rm E}$ parameter space exhibit a band-like structure, whose location and extent vary with progenitor type.

As illustrated in Fig.~\ref{fig:ry} and Fig.~\ref{fig:bw}, for a given $M_{\rm BH}$, the peak temperatures $T_{\rm peak}$ and visibility timescales $t_{\rm vis}$ of disk-SNe are generally of the same order of magnitude across different progenitor types. To identify observational signatures that are robust against progenitor uncertainties, we consider the intersection of the distinguishable regions for all four progenitor types, requiring $t_{\rm vis}>0$ and excluding the grey regions. Since these common regions cannot be reproduced by isolated SNe within the parameter space explored in this work, the corresponding combinations of $T_{\rm peak}$ and $t_{\rm vis}$ constitute distinctive observational signatures of disk SNe. The resulting ranges of $T_{\rm peak}$ and $t_{\rm vis}$ therefore provide potential diagnostics for identifying disk SNe across different black hole masses.

For $M_{\rm BH}=10^6M_{\odot}$, we find $T_{\rm peak}\sim50$--$90$ eV, corresponding to the EUV band, while the visibility timescale is relatively short, with $t_{\rm vis}\sim0.5$--$1.5$ days. For $M_{\rm BH}=10^7M_{\odot}$, the characteristic peak temperature shifts slightly toward lower energies, with $T_{\rm peak}\sim30$--$50$ eV, while the visibility timescale increases to $t_{\rm vis}\sim0.2$--$7$ days. For $M_{\rm BH}=10^8M_{\odot}$, although a large fraction of the parameter space is distinguishable from isolated SNe, only a limited portion of this region has $t_{\rm vis}>0$. In this common region, we find $T_{\rm peak}\sim3$--$8$ eV, corresponding primarily to the UV band, with a much broader range of visibility timescales, $t_{\rm vis}\sim3$--$70$ days.

Overall, these results indicate a clear dependence of the distinctive observational signatures on the black hole mass. For AGNs with $M_{\rm BH}\lesssim10^7M_{\odot}$, the unique signatures of disk SNe are concentrated primarily in the EUV band, but their visibility timescales are relatively short, typically lasting no more than several days. In contrast, for $M_{\rm BH}\sim10^8M_{\odot}$, the distinctive signatures shift toward the UV and optical bands, which are more readily accessible to observations, and can persist for several tens of days. Thus, disk SNe in lower-mass AGNs may require sensitive and rapid EUV follow-up observations, whereas those in more massive AGNs may offer a more favorable observational window through their longer-lived UV/optical emission.

\section{Summary and Discussions}
\label{sec:sum}
It has been proposed that a substantial population of stars is embedded in AGN accretion disks, where they can continue to evolve and eventually explode as SNe. The detection of such disk SNe would provide direct observational evidence for such a picture. Although previous studies have explored the physical processes and light curve evolution of SNe in AGN disks, their electromagnetic observability and characteristic observational signatures have received comparatively less systematic investigation.

In this work, we investigate the electromagnetic emission from SNe occurring in AGN disks. We construct an analytic model to describe the dynamics of the shock formed by the interaction between the SN ejecta and the surrounding disk material. To capture both light curve evolution and characteristic photon energies, we develop a one-zone radiation model that describes the evolution of the radiation energy, photon number, and electron energy in the unshocked region. We apply this framework to four representative progenitor types—red supergiants, yellow supergiants, blue supergiants, and Wolf-Rayet stars- across a range of AGN BH masses and Eddington ratios to investigate their observational signatures. We find that SNe occurring in AGN disks around lower-mass BHs or at lower Eddington ratios are generally more readily distinguishable from the underlying AGN emission and can remain detectable over a broader range of electromagnetic bands, extending from the infrared to soft X-rays.

However, the physical similarities between disk-SNe and isolated CSM-SNe can lead to degeneracies in their observational signatures, making it difficult to distinguish between them even after detection. To investigate this issue, we perform a systematic comparison between disk-SNe and isolated SNe, both with and without CSM interaction, over a broad parameter space. By statistically comparing their peak luminosities \(L_{\rm peak}\), the radiation temperature at the time of peak luminosity  \(T_{\rm peak}\), and characteristic visibility timescales \(t_{\rm vis}\), we find that a substantial fraction of disk SNe exhibit observational properties that are degenerate with those of isolated SNe. The extent of this degeneracy varies with the progenitor type and black hole mass.

Despite these degeneracies, we identify distinct observational signatures of disk SNe that remain robust across all four progenitor types for each black hole mass considered. For \(M_{\rm BH}=10^6\,M_\odot\), disk SNe exhibit a short visibility timescale of \(\lesssim 1\) day in the EUV band, with characteristic peak temperatures of \(T_{\rm peak}\sim 50\text{--}90\) eV. For \(M_{\rm BH}=10^7\,M_\odot\), the characteristic emission shifts toward lower photon energies while remaining in the EUV band, and the visibility timescale extends to several days. For \(M_{\rm BH}=10^8\,M_\odot\), the emission shifts further toward the UV and optical bands, with visibility timescales reaching tens of days.

We emphasize that the observational degeneracies identified in this work are based on a comparison of a limited set of macroscopic observables, namely \(L_{\rm peak}\), \(T_{\rm peak}\), and \(t_{\rm vis}\). This does not necessarily imply that disk SNe are indistinguishable from isolated SNe in all observational aspects. In particular, differences in the CSM density structure, geometry, and chemical composition can affect the shock dynamics, ionization state, and radiative transfer, giving rise to distinct spectral features and temporal or multi-wavelength signatures \citep{fransson2002optical,chevalier2003supernova,fraser2020supernovae,dessart2023using}. Detailed spectroscopy can therefore provide information on the CSM properties that is not captured by the macroscopic observables considered in our analysis. A more comprehensive investigation incorporating detailed spectral modeling and additional observational diagnostics is therefore warranted to identify potentially unique signatures of SNe occurring in AGN disks.

Our results have important implications for future observations of disk SNe. For \(M_{\rm BH}=10^6\,M_\odot\), the distinctive emission primarily lies in the EUV band and lasts for less than a day, posing significant challenges for direct detection because of the limited availability of wide-field EUV facilities, strong interstellar absorption, and the short visibility timescale. For \(M_{\rm BH}=10^7\,M_\odot\), the distinctive emission remains primarily in the EUV but persists for several days, providing a relatively longer window for follow-up observations, although absorption remains an important limitation. In contrast, for \(M_{\rm BH}=10^8\,M_\odot\), the distinctive emission shifts toward the UV and optical bands and remains visible for tens of days, making these events more accessible to existing and upcoming time-domain facilities. Existing UV observations with the Neil Gehrels Swift Observatory/UVOT have demonstrated the diagnostic power of UV light curves for supernovae \citep{brown2009ultraviolet}, while HST UV spectroscopy has revealed signatures of circumstellar interaction that are not readily captured by broadband photometry alone \citep{ben2015ultraviolet,bostroem2024circumstellar}. Looking ahead, the Ultraviolet Transient Astronomy Satellite (ULTRASAT), with its wide field of view and near-UV sensitivity, is particularly promising for discovering UV-bright transients on timescales relevant to our predictions \citep{shvartzvald2024ultrasat}. The proposed Ultraviolet Explorer (UVEX), combining wide-field UV imaging with spectroscopic follow-up, may provide additional capabilities for characterizing short-lived UV transients \citep{kulkarni2021science}. In the optical band, the Vera C. Rubin Observatory Legacy Survey of Space and Time (LSST) will provide high-cadence, wide-field monitoring of the transient sky \citep{ivezic2019lsst}. Coordinated UV and optical observations with these facilities, together with archival AGN light curves and rapid follow-up observations, may therefore provide a promising strategy for identifying disk SNe and distinguishing them from isolated SNe with or without CSM interaction.

From the perspective of multi-messenger astrophysics, explosion events including SNe in AGN disks may also produce high-energy neutrinos through hadronic interactions between shock-accelerated protons and the dense disk material~\citep{zhu2021high,zhou2023high,zhou2023high2,ma2024high}. These neutrinos may be searched for by current and future neutrino observatories, including IceCube~\citep{aartsen2018astrophysical}, KM3NeT/ARCA~\citep{km3net2024astronomy}, and IceCube-Gen2~\citep{clark2021icecube}. The detection of a temporally and spatially coincident neutrino signal, together with an electromagnetic transient exhibiting the characteristic properties predicted in this work, would provide a powerful multi-messenger approach to identifying SNe occurring in AGN disks.

\bibliography{Reference}{}
\bibliographystyle{aasjournalv7}



\end{document}